\documentclass{article}
\usepackage[utf8]{inputenc}
\usepackage{geometry}
\usepackage{subcaption}
\usepackage{hyperref, url}
\usepackage{lineno}
\usepackage{setspace}
\usepackage{authblk}
\usepackage{orcidlink}
\usepackage[style=numeric-comp,sorting=none]{biblatex}
\usepackage{amsmath}

\title{Evaluating local climate in global storm-resolving models with the K\"{o}ppen-Geiger classification}

\author[1]{Chiel C. van Heerwaarden\,\orcidlink{0000-0001-7202-3525} \texttt{(chiel.vanheerwaarden@wur.nl)}}
\author[1]{Menno A. Veerman\,\orcidlink{0000-0002-4869-3948}}

\author[1]{Imme Benedict\,\orcidlink{0000-0002-1946-6332}}
\author[2]{Lukas Brunner\,\orcidlink{0000-0001-5760-4524}}
\author[3,4]{Edgar Dolores-Tesillos\,\orcidlink{0009-0005-7327-180X}}
\author[5]{Emanuel Dutra\,\orcidlink{0000-0002-0643-2643}}
\author[6]{Erich Fischer\,\orcidlink{0000-0003-1931-6737}}
\author[7]{Junhong Lee\,\orcidlink{0000-0001-6459-9284}}
\author[4]{Olivia Martius\,\orcidlink{0000-0002-8645-4702}}
\author[8]{Xabier Pedruzo-Bagazgoitia\,\orcidlink{0000-0001-5129-6364}}
\author[9]{Ulrike Proske\,\orcidlink{0000-0002-3153-7228}}
\author[1,10]{Sarah N. Warnau\,\orcidlink{0000-0002-2535-6789}}
\author[6,11]{Jonathan D. Wille\,\orcidlink{0000-0002-3918-5204}}

\author[12]{Cathy Hohenegger\,\orcidlink{0000-0002-7478-6275}}

\affil[1]{Meteorology and Air Quality Group, Wageningen University \& Research, Wageningen, The Netherlands}
\affil[2]{ Research Unit for Sustainability and Climate Risks, Earth and Society Research Hub (ESRAH), University of Hamburg, Hamburg, Germany}
\affil[3]{Faculty of Geosciences and Environment, University of Lausanne, Lausanne, Switzerland}
\affil[4]{Institute of Geography and Oeschger Centre for Climate Change Research, University of Bern, Bern, Switzerland}
\affil[5]{Instituto Português do Mar e Atmosfera, Lisbon, Portugal}
\affil[6]{Institute for Atmospheric and Climate Science, ETH Zurich, Zurich, Switzerland}
\affil[7]{Climate Prediction Research Center, Seoul National University, Seoul, Republic of Korea}
\affil[8]{European Centre for Medium-Range Weather Forecasts (ECMWF), Bonn, Germany}
\affil[9]{Hydrology and Environmental Hydraulics Group, Wageningen University \& Research, Wageningen, The Netherlands}
\affil[10]{Wetsus, Leeuwarden, The Netherlands}
\affil[11]{Biogéosciences, CNRS/Université Bourgogne Europe, Dijon, France}
\affil[12]{Max Planck Institute for Meteorology, Hamburg, Germany}

\date{\today}

\usepackage{graphicx}

\begin{document}

\maketitle

\begin{abstract}
Global storm-resolving models aspire to become digital twins of the Earth, delivering information at the local scale at which humans experience climate. We evaluated how well two such models, ICON and IFS-FESOM, reproduce the climate as classified by the K\"{o}ppen-Geiger system, using 30-year (2020--2049) simulations from the nextGEMS project at approximately 9~km global resolution under the SSP3-7.0 scenario.

Both models capture the global distribution of the five main climate categories, encouraging given the infancy of coupled storm-resolving climate modelling. Substantial regional biases nonetheless remain. Both underestimate tropical rainforest (Af) extent due to insufficient dry-month precipitation in Amazonia and equatorial Africa. ICON almost eliminates hot arid desert (BWh) across Australia through excessive precipitation, while IFS-FESOM reproduces it well. The two models also show opposing biases along the temperate--continental boundary: IFS-FESOM winters are too cold in western Europe, ICON winters too warm. Substituting observed temperature or precipitation into the model fields reveals that precipitation errors dominate misclassification, while temperature biases play a secondary role confined to mid-latitude climate zone boundaries.

Under climate change, the two models and the CMIP6 projections agree on the direction of climate zone shifts: expansion of tropical savanna and hot desert at the expense of subarctic, tundra, and ice cap zones. However, inter-model differences in present-day climate exceed the 30-year climate change signal for many zones, calling for caution in regional projections and adaptation planning. Our results expose where local-scale climate representation still falls short of the digital twin ambition, while confirming that storm-resolving models already perform well across many regions. We propose K\"{o}ppen-Geiger classification as a standard diagnostic to help track further progress.
\end{abstract}

\section{Introduction}
Many modeling centers around the world have taken up the challenge of building \textit{storm-resolving} earth system models \cite{Satoh2019, Stevens2019, Donahue2024}. These run at horizontal grid spacings approaching the kilometer scale and resolve meso-scale phenomena such as deep-convective precipitation systems, landscape-driven circulations, and ocean meso-scale eddies.

Pioneering studies with coupled storm-resolving models already demonstrate improvements in the representation of local atmospheric phenomena, for instance in the soil moisture--precipitation feedback \cite{Lee2024, Segura2025, Lee2026}, extreme precipitation intensities \cite{Wille2025}, the evolution of maritime stratocumulus fields \cite{Segura2025}, the impact of sea ice on the atmosphere \cite{Rackow2024}, or the representation of the surface radiation balance \cite{Veerman2026, XabiUrban2025} and various climate extremes \cite{Brunner2025} over complex terrain. Large-scale circulation also improves, although challenges remain in representing atmospheric blocking \cite{DoloresTesillos2025StormResolving}.

The ambition is for these models to become \textit{digital twins} of the earth system that deliver not only global consistency, but also local information at a resolution relevant for decision making. The European Destination Earth (DestinE) programme, for instance, foresees their interactive use as key tools in the management of renewable energy, water, and food \cite{DoblasReyes2025}.

Storm-resolving models thus bring back the human dimension in climate modelling.
In the 19th century bottom-up view of climate, pioneered by Alexander von Humboldt, local experience was key and global climate and understanding thereof emerged as the aggregate of local observations \cite{Humboldt1817, heymannEvolutionClimateIdeas2010}.
Humboldt's conception of climate was adopted by pioneers such as Hann and K\"{o}ppen \cite{Hann1883, Koppen1884} and gave rise to the field of classical climatology, where building an accurate picture of atmospheric phenomena at a given location was the central task. One of the most influential products of this era is the still widely used K\"{o}ppen-Geiger climate classification system \cite{Peel2007, Beck2018, Beck2023}, which aggregates local climate into climate zones based on monthly mean air temperature and monthly precipitation sums.

However, 20th-century climate science, and earth system modelling in particular, has prioritized global scales, with a strong focus on the global energy balance at the top of the atmosphere.
This top-down approach has been remarkably successful: it revealed the interplay of the atmosphere and ocean in shaping climate \cite{ManabeBryan1969} and enabled the first quantitative predictions of the climate impact of rising atmospheric CO$_2$ \cite{ManabeWetherald1967, Hansen1981} concentrations.
Yet this success came at the expense of local scales -- the ones that matter most to human experience \cite{heymannClimateChangeDilemma2019}.
Regional modelling initiatives such as CORDEX \cite{Gutowski2016} have addressed this gap, but feedbacks from regional to global scales remain absent.
Storm-resolving global earth system models now offer the possibility to come full circle, as they promise to provide ``local granularity, globally'' \cite{stevensEarthVirtualizationEngines2024}.

In our view, global storm-resolving models that aim to be digital twins should capture climate from the bottom up, following Humboldt's philosophy, with an accurate representation of local and regional climate. In this study, performed within the \textit{Storms and Land} theme of the European H2020 Next Generation Earth Modelling Systems (nextGEMS) project, we evaluate how well two such models, IFS-FESOM (ECMWF) and ICON (MPI), represent local climate over land, using 30-year (2020--2049) simulations at 9~km global horizontal resolution. Standing on the shoulders of the 19th-century pioneers of local climate classification, we apply the K\"{o}ppen-Geiger classification to model output and assess how well the models reproduce the global distribution of local climate zones derived from observations \cite{Beck2018}. We also compare the two models in terms of climate zone migration under climate change.

\section{Methods}
\subsection{Kilometer-scale Earth System Models}
We used the multi-year global climate simulations of the fourth \cite{Wieners2024} cycle of the nextGEMS project \cite{Segura2025}.
The IFS-FESOM simulations are based on IFS-FESOM version 48r1 \cite{ECMWF2023} coupled to the FESOM2.5 \cite{Rackow2023} ocean model, see \cite{Rackow2024} for a more detailed description.
The ICON simulation is based on the configuration described in \cite{Hohenegger2023}, with modifications described by \cite{Segura2025}. This version of ICON is designed specifically for (sub)km-scale simulations and contains fewer parametrizations than IFS-FESOM. 
 
Both models were integrated for 30 years under the SSP3-7.0 scenario \cite{Oneill2016} with time-varying greenhouse gas and ozone concentrations.
ICON started 1 January 2020 at approximately 10~km atmospheric and 5~km oceanic resolution; IFS-FESOM started 20 January 2020 at approximately 9~km atmospheric and 5~km oceanic resolution.
Output was post-processed onto a Hierarchical Equal Area IsoLatitude Pixelation (HEALPix) grid \cite{Gorski2005} at $\sim$12.7~km horizontal resolution.

\subsection{K\"{o}ppen-Geiger climate classification}
We use the K\"{o}ppen-Geiger classification as described in \cite{Peel2007}. It takes monthly mean air temperature ($^{\circ}$C) and monthly precipitation sums (mm) as input and divides the world into tropical (A), desert (B), temperate (C), continental (D), and polar (E) climates, with subdivisions based on temperature and precipitation thresholds. The full decision tree is given in Supplement A, Fig.~\ref{fig:koppen_decision_tree}.

As observational reference we use the high-resolution (1~km) K\"{o}ppen-Geiger maps of \cite{Beck2018}, updated by \cite{Beck2023}, which are based on 30-year climatologies of gauge-based temperature (CRU~TS, 0.5$^{\circ}$) and merged satellite-gauge precipitation (CHPclim, 0.05$^{\circ}$) data, downscaled to 1~km using high-resolution auxiliary fields.
We compare the present-day IFS-FESOM and ICON model climate (2021--2025) against the 1991--2020 observational period. For future climate, \cite{Beck2023} additionally provide K\"{o}ppen-Geiger projections derived from downscaled and bias-corrected CMIP6 ensemble output, which serve as an independent reference for comparison against the IFS-FESOM and ICON climate (2045--2049).

\section{Results}

\subsection{Present-day climate classification}

\begin{figure}
    \centering
    \includegraphics[width=0.95\linewidth]{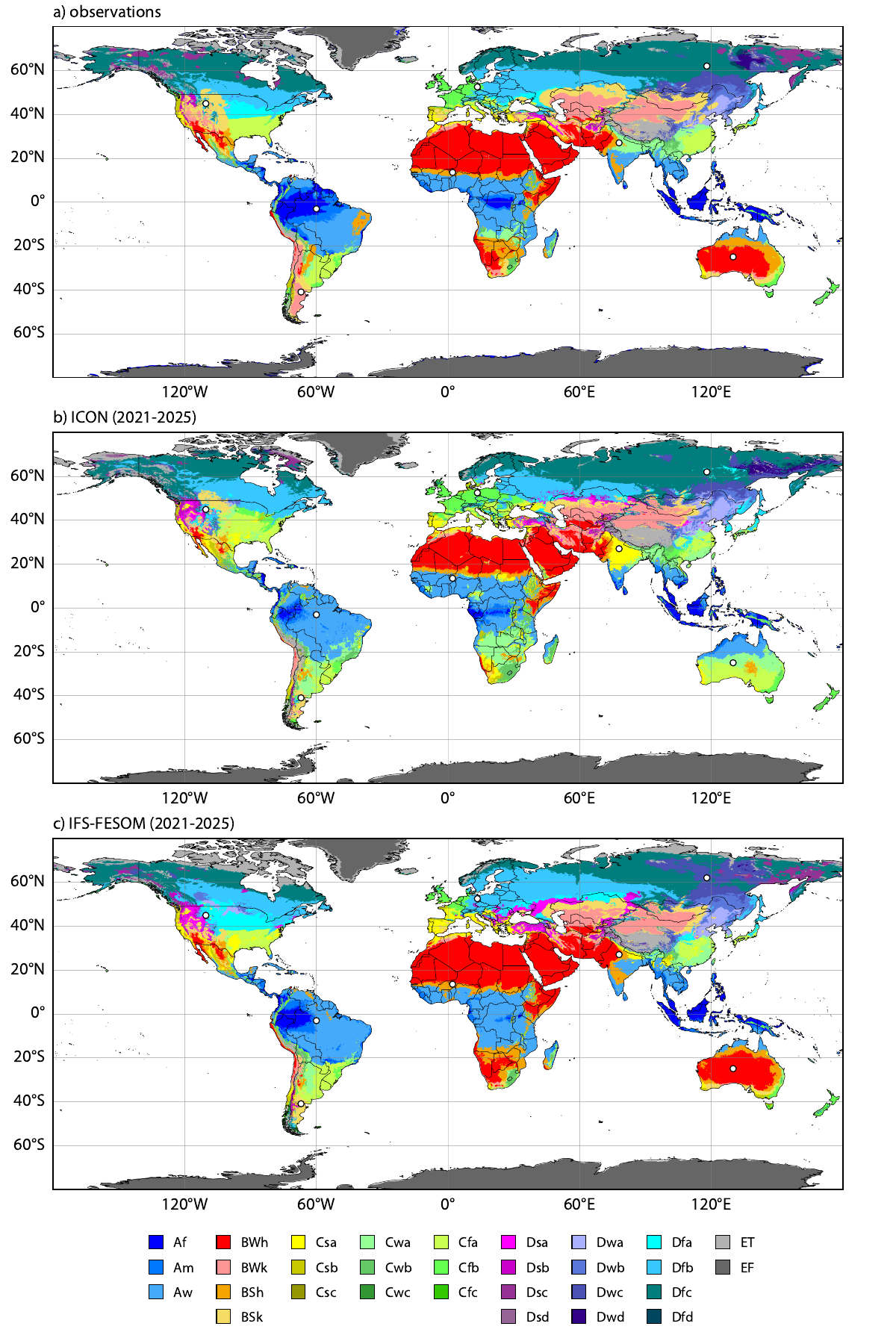}
    \caption{\textbf{K\"{o}ppen-Geiger climate classification from observations and storm-resolving models.} Dominant climate zone for \textbf{a)} the observational reference of Beck et al. (2018), \textbf{b)} ICON, and \textbf{c)} IFS-FESOM, for the period 2021--2025. Major climate zones are tropical (A), arid (B), temperate (C), continental (D), and polar (E), see Supplement A for a full overview. Markers indicate the eight locations examined in Sec.~\ref{sec:location_panel}.}
    \label{fig:koppen_present_panels}
\end{figure}

Figure \ref{fig:koppen_present_panels} compares the K\"{o}ppen-Geiger classification derived from ICON and IFS-FESOM output with the reference of \cite{Beck2018}. At a global scale, both models capture the large-scale patterns of temperature and precipitation, reproducing the broad distribution of the five main climate categories: the tropical belt across South America, Africa, and Southeast Asia; the arid regions of the Sahara, Arabian Peninsula, and Central Asia; the temperate mid-latitudes; the continental climates of northern Eurasia and North America; and the polar climates at high latitudes and on the Tibetan Plateau.

Closer inspection, however, reveals substantial regional differences -- between the models and the reference, and between the two models. In the tropics, both ICON and IFS-FESOM underestimate the extent of tropical rain forest (Af) climate. In Amazonia, both models lack rain forest climate near the coast and replace it with tropical monsoon (Am) or savanna (Aw) climates, indicating that precipitation in the driest months is too low. In equatorial Africa the underestimation is more pronounced: ICON retains very little Af in the Congo Basin, IFS-FESOM essentially none. The Maritime Continent is better represented, with both models broadly capturing tropical climates in Indonesia and Malaysia.

Inter-model differences are most striking in the arid regions. Both models reproduce the Sahara--Arabian desert belt well, but ICON almost completely lacks hot arid desert (BWh) in Australia, replacing it with temperate (Cw), tropical savanna (Aw), and semi-arid (BSh) climates. IFS-FESOM maintains the Australian desert, in closer agreement with the reference. ICON also underestimates aridity at the tip of southern Africa, pointing to a fundamental difference in the representation of the hydrological cycle over the southern hemisphere continents.

The temperate--continental (C--D) boundary reveals another systematic inter-model difference. In IFS-FESOM, continental climates spread too far west into Europe, suggesting winters that are too cold. ICON shows the opposite tendency: temperate climates extend too far into the continental interior, both in Europe and the Midwestern United States, where the reference indicates continental (Dfa/Dfb) climates. The two models thus have opposing biases along this boundary, pointing to different representations of winter temperature climatology.

In India, both models differ from the reference in the distribution of tropical and arid subcategories (A, B and C patterns), reflecting the challenge of representing monsoon precipitation and its sharp spatial gradients. In West Africa, the Sahelian transition from desert to savanna is broadly captured, but differences in the position and intensity of the monsoon rain belt affect the placement of the BSh--Aw boundary.

While the models agree with the reference on the global pattern, they disagree with each other in several key regions. To understand the origin of these misclassifications, we examine the monthly temperature and precipitation cycles at eight representative locations.

\subsection{Monthly climate in eight regions of interest} \label{sec:location_panel}
\begin{figure}
    \centering
    \includegraphics[width=0.95\linewidth]{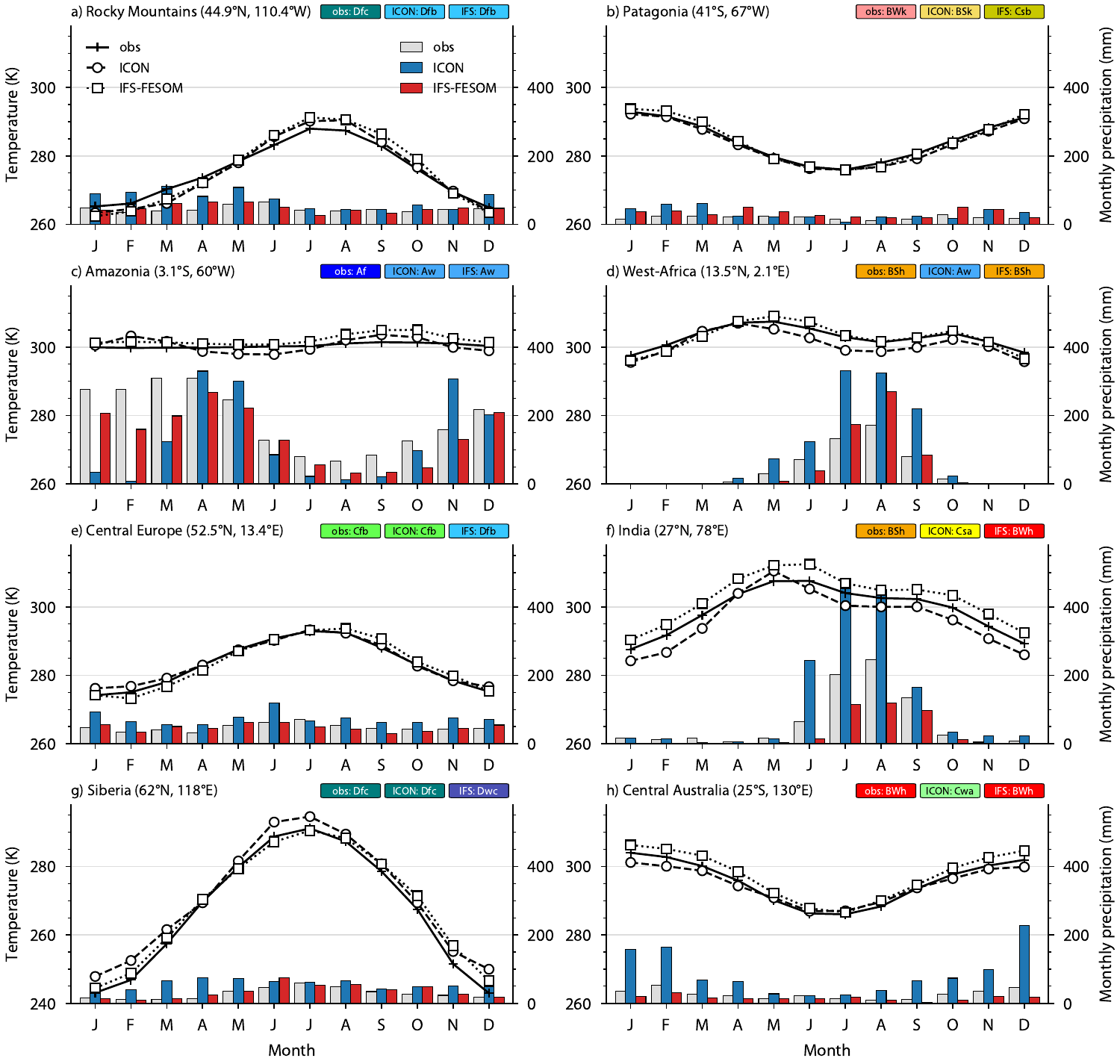}
    \caption{\textbf{Monthly climatology of temperature and precipitation at eight locations with classification discrepancies.} Annual cycle of monthly mean 2-metre temperature (lines, left axis) and precipitation (bars, right axis) for observations (gray), ICON (blue), and IFS-FESOM (red) over the period 2021--2025. Locations are indicated in Fig.~\ref{fig:koppen_present_panels}a, and are ordered from west to east.}
    \label{fig:location_panel}
\end{figure}

Figure \ref{fig:location_panel} shows the monthly mean temperature and precipitation at these eight locations for IFS-FESOM, ICON, and the observations from \cite{Beck2023}.

\subsubsection{Rocky Mountains (44.9$^{\circ}$N, 110.4$^{\circ}$W)}
The Rocky Mountains are challenging due to steep gradients in altitude, temperature, and precipitation. Both models share a similar temperature cycle, with summers roughly 5~K too warm and winters 3~K too cold (Figure \ref{fig:location_panel}a). The summer warm bias pushes more than three months above the 10$^{\circ}$C threshold, shifting the classification from subarctic continental (Dfc) to warm-summer continental (Dfb), consistent with the northward expansion of temperate climates in Figure \ref{fig:koppen_present_panels}. ICON substantially overestimates winter and spring precipitation, whereas IFS-FESOM tracks the observations year-round.

\subsubsection{Patagonia (41$^{\circ}$S, 67$^{\circ}$W)}
In Patagonia, both models reproduce the observed temperature cycle well (Figure \ref{fig:location_panel}b). Discrepancies emerge in precipitation: both exaggerate its seasonal cycle, whereas observations are relatively uniform year-round. ICON overestimates most strongly during austral summer, while IFS-FESOM is overall slightly too wet across all seasons. These small biases flip the classification between temperate (C) and cold arid (BW) zones, even though the broad climate is captured.

\subsubsection{Amazonia (3.1$^{\circ}$S, 60$^{\circ}$W)}
In Amazonia, both models represent temperature well year-round (Figure \ref{fig:location_panel}c). The discrepancy lies in precipitation: observations exceed 200~mm each month through DJF and MAM, peaking in early spring. Both models underestimate this, but differently: ICON develops a pronounced January--March dry season largely absent in the observations, whereas IFS-FESOM is consistently drier year-round without such a seasonal contrast. Either way, the driest-month threshold is breached and the rain forest (Af) classification is lost. The two models thus reach the same misclassification for different reasons: an erroneous dry season in ICON, a general precipitation deficit in IFS-FESOM.

\subsubsection{West-Africa (13.5$^{\circ}$N, 2.1$^{\circ}$E)}
Both models capture the seasonal temperature cycle at Niamey well (Figure \ref{fig:location_panel}d), including the characteristic dip when monsoon cloud cover and wet soils suppress surface warming; in ICON this dip is about 3~K too strong. Both simulate the monsoon precipitation signal, with onset around June and retreat around October, but overestimate peak rainfall. ICON is most extreme, reaching roughly 300~mm in July--August versus 200~mm observed; the associated cloud cover and soil moisture likely drive its temperature underestimation. IFS-FESOM stays closer to the observed totals, preserving the hot semi-arid (BSh) classification, whereas ICON's wet bias shifts it to tropical savanna (Aw).

\subsubsection{Central Europe (52.5$^{\circ}$N, 13.4$^{\circ}$E)}
In Central Europe (Figure \ref{fig:location_panel}e), IFS-FESOM has a slight cold bias in the first half of the year and a slight warm bias afterward, while ICON runs several degrees too warm in DJF. The cold bias pushes IFS-FESOM towards continental (D), whereas both the reference and ICON remain temperate (C). Spatially, ICON extends temperate climates eastward to the Poland--Belarus border, while IFS-FESOM pushes continental climates westward, almost to the German--Dutch border. Precipitation is relatively uniform in all three datasets, though consistently too high in ICON; still, classification errors here are temperature-driven.

\subsubsection{India (27$^{\circ}$N, 78$^{\circ}$E)}
At this Indian monsoon location (Figure \ref{fig:location_panel}f), the two models show mirror-image biases. IFS-FESOM runs about 5~K too warm year-round, ICON about 5~K too cold. Precipitation reverses: ICON doubles JJA monsoon rainfall, while IFS-FESOM underestimates it. The two may be physically linked through evaporative cooling and cloud cover: insufficient monsoon precipitation permits IFS-FESOM's warm bias, and vice versa for ICON. Both are misclassified: observations indicate hot semi-arid steppe (BSh), but ICON produces temperate with hot, dry summers (Csa) from excess precipitation, while IFS-FESOM yields hot arid desert (BWh) from its monsoon deficit.

\subsubsection{Siberia (62$^{\circ}$N, 118$^{\circ}$E)}\label{sec:location_panel_siberia}
In Siberia (Figure \ref{fig:location_panel}g), both models capture the enormous annual temperature cycle of approximately 45~K. Counterintuitively, IFS-FESOM is misclassified despite smaller precipitation biases, while ICON retains the correct class despite large winter precipitation errors. The K\"{o}ppen precipitation thresholds between continental subcategories depend on the annual temperature distribution, so a small temperature bias can shift the threshold enough to flip the classification. A correct K\"{o}ppen-Geiger classification therefore does not necessarily imply a more accurate simulation.

\subsubsection{Central Australia (25$^{\circ}$S, 130$^{\circ}$E)}\label{sec:location_panel_australia}
Central Australia (Figure \ref{fig:location_panel}h) is one of the starkest cases of model disagreement. IFS-FESOM runs too warm but reproduces the observed low precipitation, preserving the arid desert (BW) classification. ICON, by contrast, produces fourfold more precipitation than observed during the austral summer (DJF), erasing the dry conditions that define a desert climate. This excess explains the complete absence of hot arid desert (BWh) across ICON's Australia (Figure \ref{fig:koppen_present_panels}) and represents a fundamental misrepresentation of the regional water cycle.

\subsection{Tracing the reason of mismatch}

\begin{figure}
    \centering
    \includegraphics[width=0.95\linewidth]{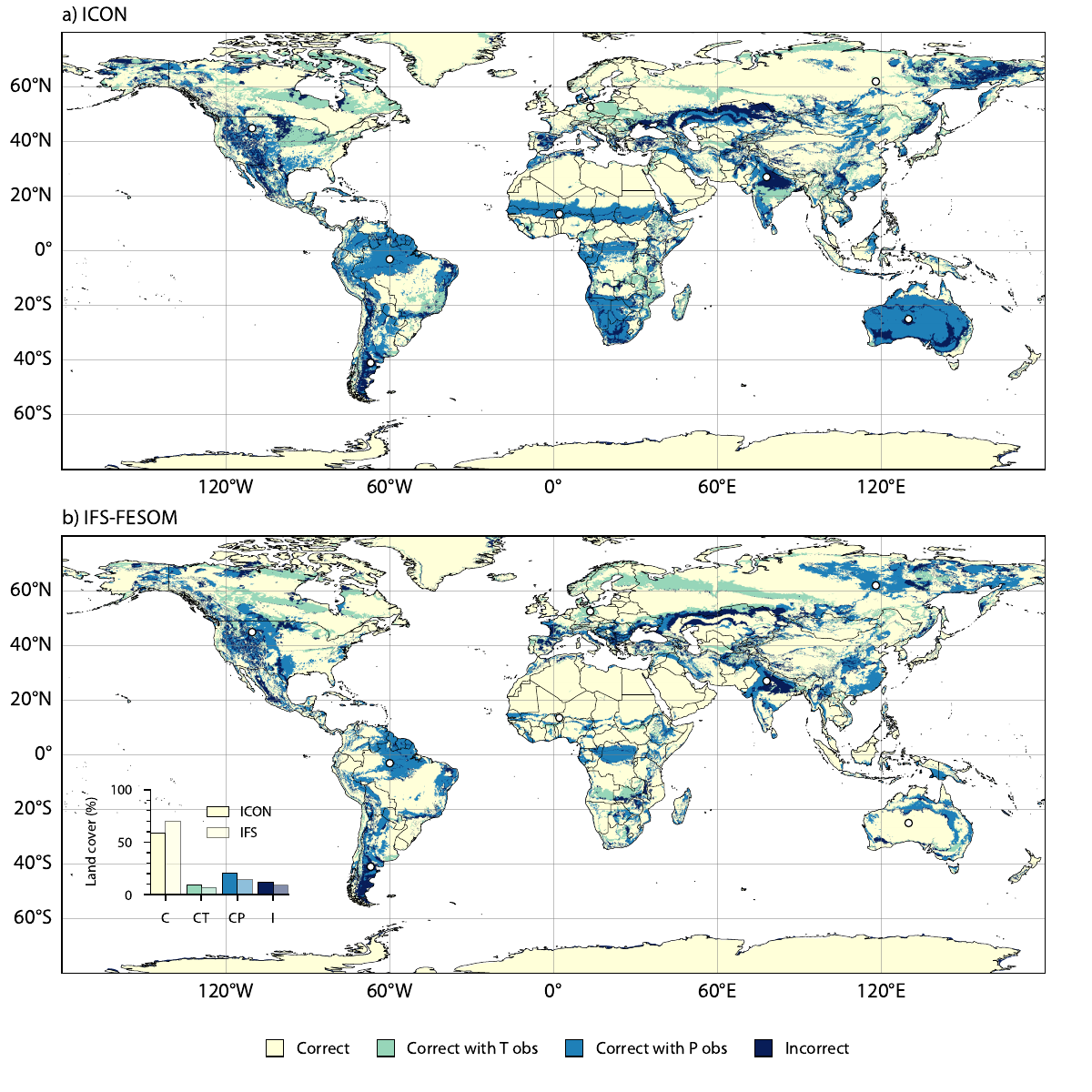}
    \caption{\textbf{Attribution of climate classification errors to temperature and precipitation biases.} The K\"{o}ppen-Geiger classification is recomputed for \textbf{a)} ICON and \textbf{b)} IFS-FESOM after substituting either the model's temperature or precipitation with the corresponding observed field from Beck et al. (2018). Grid points are coloured according to whether the correct classification is restored by replacing temperature (green), precipitation (blue), or neither variable alone (dark blue). Correctly classified points are shown in yellow.}
    \label{fig:koppen_present_cross_panels}
\end{figure}

To systematically attribute misclassification across the globe, we recompute the K\"{o}ppen-Geiger classification after substituting either the model's temperature or precipitation with the corresponding field from the Beck reference dataset (Figure \ref{fig:koppen_present_cross_panels}). If replacing temperature corrects the classification we attribute the error to temperature (``Correct with T obs''); if replacing precipitation does, we attribute it to precipitation (``Correct with P obs''). Grid points where neither substitution alone restores the correct class are labeled ``Incorrect'', meaning both variables need to be corrected simultaneously.

Summed over the globe (inset in Figure \ref{fig:koppen_present_cross_panels}), approximately two-thirds of all grid points are classified correctly, with IFS-FESOM outperforming ICON by roughly ten percentage points. Among the misclassified grid points, precipitation errors account for the largest share in both models.

Blue shading (precipitation-driven errors) is far more widespread than green (temperature-driven). In ICON, precipitation drives misclassification across the tropics (Amazonia, equatorial Africa, the Indian subcontinent) and nearly the entire Australian continent. IFS-FESOM shows similar precipitation-driven tropical errors (Amazonia, central Africa, India, the Maritime Continent) but over a smaller extent than ICON. 

The contrast between the two models is most pronounced in the southern hemisphere. In ICON, Australia appears almost entirely blue, confirming that the excessive precipitation identified in Section \ref{sec:location_panel_australia} is not a local anomaly but a continent-wide bias. IFS-FESOM correctly classifies most of Australia, consistent with its more realistic representation of aridity. A similar pattern appears in southern Africa, where ICON shows substantially more precipitation-driven errors than IFS-FESOM.

Temperature-driven errors are secondary in spatial extent but geographically systematic. They appear primarily along the temperate--continental (C--D) boundary: in western North America, where ICON's warm bias shifts classification towards temperate climates; across northern Europe and Russia, where winter temperature biases affect the boundary placement; and in parts of central Asia. These patterns confirm that mid-latitude climate zone misclassifications are predominantly temperature-driven.

Regions where neither substitution alone corrects the classification are limited but appear in parts of southern South America, scattered locations in Africa, and some mountainous areas, where both variables are simultaneously biased.

In summary, precipitation is the dominant source of misclassification in both models, with ICON more affected than IFS-FESOM. Temperature biases play a secondary but systematic role, confined to mid-latitude climate zone boundaries. Improving the representation of tropical and subtropical precipitation should therefore be a priority for advancing the fidelity of these storm-resolving models. 

\subsection{Climate change over 30 years}

\begin{figure}
    \centering
    \includegraphics[width=0.95\linewidth]{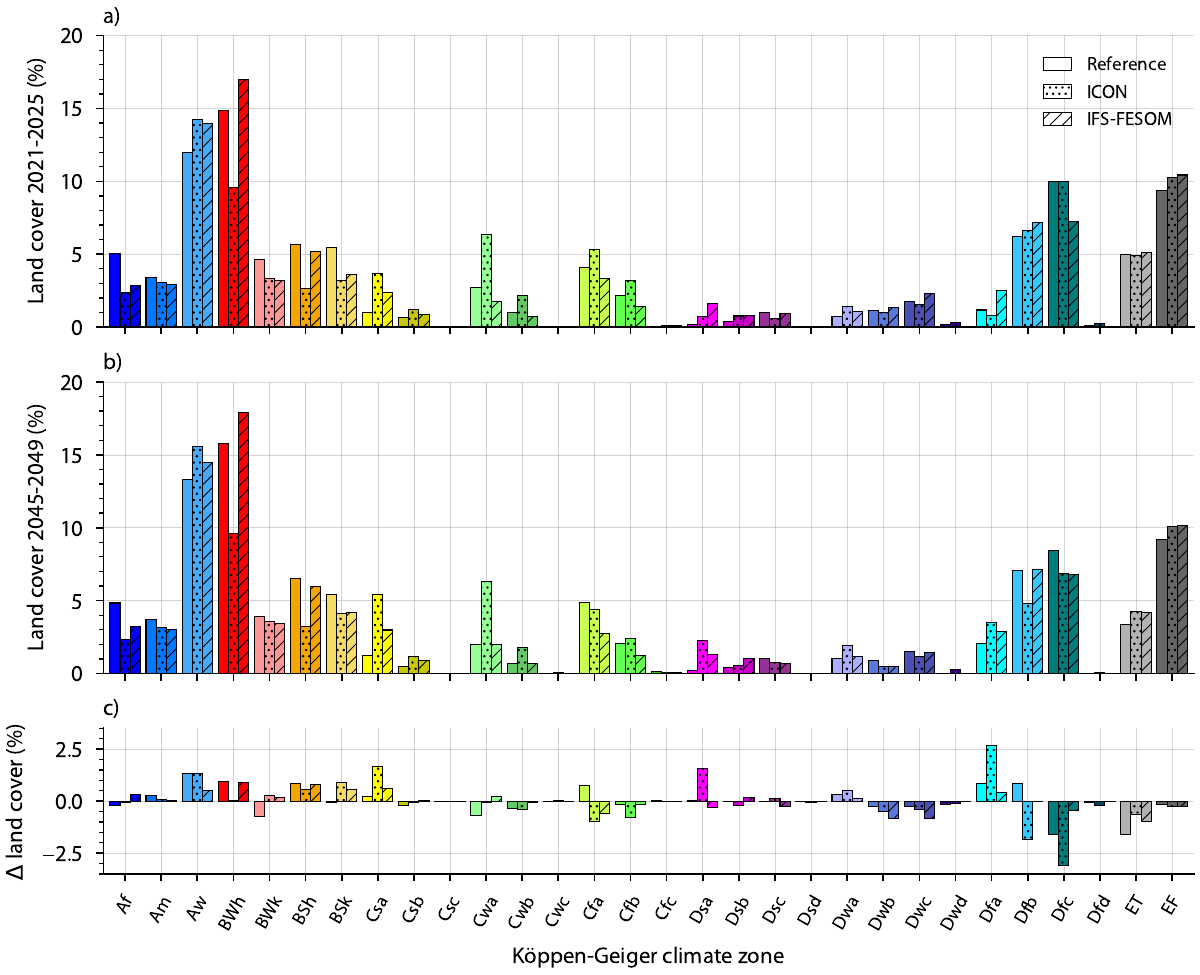}
    \caption{\textbf{Present-day and future land cover fractions by K\"{o}ppen-Geiger climate zones.} Percentage of global land area occupied by each climate zone for the observational reference (no hatch); Beck et al. 2018 for present, CMIP6-based projections from Beck et al. 2023 for future), ICON (dotted) and IFS-FESOM (diagonal stripes). \textbf{a)} present-day period (2021--2025). \textbf{b)} future period (2045--2049) under the SSP3-7.0 scenario. \textbf{c)} difference between future and present, showing the climate change signal.}
    \label{fig:koppen_bar_charts}
\end{figure}

The 30-year simulations under SSP3-7.0 allow us to examine how climate zones shift over time and to compare the climate change signal against inter-model spread and CMIP6 projections. Figure \ref{fig:koppen_bar_charts} shows the land area fraction occupied by each K\"{o}ppen-Geiger climate zone for the present (2021--2025) and future (2045--2049) periods, as well as the difference between the two.

The top panel allows for estimation of biases in terms of global land cover statistics. The most prominent discrepancy is in tropical rain forest (Af), where both models underestimate the reference land fraction, and in tropical savanna (Aw), where both overestimate it. This is consistent with the loss of rain forest classification due to insufficient dry-month precipitation. In the arid zones, ICON's deficit in hot desert (BWh) stands out, with approximately one-third less land area than the reference, while IFS-FESOM slightly overestimates BWh. Among the temperate climates, IFS-FESOM substantially underestimates humid subtropical (Cfa), reflecting its tendency to extend continental climates too far into temperate regions. For the continental and polar zones (D and E), the models broadly agree with the reference: subarctic (Dfc), tundra (ET), and ice cap (EF) are well reproduced in global area.

The middle panel shows the classification for 2045--2049. The reference bars are CMIP6 projections \cite{Beck2023}, providing an independent source. The overall pattern remains similar, but systematic shifts are visible, made explicit in the bottom panel.

Despite the relatively short 30-year window under SSP3-7.0, we find mostly clear and physically consistent trends, but also some exceptions. In all three datasets (ICON, IFS-FESOM, and the CMIP6 reference), tropical savanna (Aw) and hot desert (BWh) expand, while subarctic continental (Dfc), tundra (ET), and ice cap (EF) contract, consistent with the expected poleward expansion of tropical and arid zones at the expense of cold continental and polar climates \cite{Song2025, Bayar2023}. Both models deviate from CMIP6 by showing slight gains in cold desert (BWk), cold semi-arid steppe (BSk), and dry-winter humid subtropical (Cwa) where CMIP6 projects losses or no change. The cold desert gain suggests model warming is insufficient to shift these regions to hot desert, while the Cwa expansion points to a monsoon response that differs from the CMIP6 ensemble, allowing subtropical climates to encroach into regions CMIP6 classifies as temperate or continental. ICON specifically stands out with a large increase in hot-summer Mediterranean continental (Dsa), pointing to enhanced summer drying in continental interiors, and a strong shift from Dfc/Dfb towards hot-summer humid continental (Dfa), indicating a stronger poleward push of summer temperature thresholds than IFS-FESOM or CMIP6.

The largest signal appears in the subarctic continental zone (Dfc), which loses approximately 2 percentage points of global land area in the CMIP6 data and more than 2.5 percent in ICON, although IFS-FESOM remains largely unchanged. This contraction is partly compensated by an expansion of the warm-summer continental zone (Dfb), reflecting a northward migration of the continental climate boundary. Tundra (ET) and ice cap (EF) also shrink consistently across all three datasets.

In the tropics and arid regions, the changes are smaller in absolute terms but consistent in direction. Tropical savanna (Aw) and hot desert (BWh) both expand by roughly 1 percentage point, while the temperate zones show mixed and generally small signals. These patterns are broadly in line with CMIP6 multi-model projections \cite{Song2025, Kim2021}.

Crucially, while the direction of change is consistent across ICON, IFS-FESOM, and CMIP6, the magnitude of inter-model differences in present-day climate often exceeds the 30-year climate change signal. The difference in BWh land fraction between ICON and IFS-FESOM is approximately 7 percentage points, whereas the climate change signal in BWh is only about 1 percentage point. Similarly, the spread in temperate (Cfa) between the models is a few percentage points, while the climate change trend is near zero. Model-formulation uncertainty thus currently dominates over the forced climate change signal in many zones, though consistent changes between the two models are also observed. Regional projections and policy decisions based on these models therefore require caution.

\section{Discussion and conclusion}
Our analysis of the coupled storm-resolving models ICON and IFS-FESOM shows they capture the global distribution of K\"{o}ppen-Geiger climate zones.
Given the novelty of this modelling approach, it is encouraging that they reproduce the basic properties of the climate system.
Yet, substantial regional biases remain, particularly in tropical rainforests and arid regions, and the differences between the two models exceed the climate change signal over the 30-year simulation period.

These findings have two key implications. First, high horizontal resolution alone does not guarantee faithful representation of regional climate.
The biases we identify stem from temperature and precipitation errors that require improvements in process representation. \cite{Brunner2026} show that two versions of the km-scale IFS-FESOM model reproduce temperature far better than any coupled CMIP-class model, suggesting continued development can yield further gains in skill.
Nonetheless, we cannot rule out that even finer grids are needed.
ICON, which explicitly resolves deep convection, suffers from widespread precipitation-driven misclassification, possibly reflecting that at kilometer-scale resolution convective circulations are represented but not yet fully resolved.
Ongoing efforts to push global simulations to sub-kilometer resolution \cite[e.g.,][]{Takasuka2024} will help clarify whether further increases in resolution can reduce such biases.

Second, the large inter-model spread in present-day climate calls for long historical simulations with multiple models, as foreseen in DestinE \cite{DoblasReyes2025}, to support reliable adaptation planning, where the high resolution of storm-resolving models more readily speaks to policymakers. It is also important to study how classification errors relate to biases in ocean temperatures and the associated dynamics, as both models show large-scale dynamics closer to observations in an AMIP configuration without SST biases \cite{DoloresTesillos2025StormResolving}.

A methodological caveat is that our present-day classification is based on only five years (2021--2025) of model output, whereas the reference of \cite{Beck2018} uses 30-year climatologies. The shorter averaging period makes the model classification sensitive to interannual variability, and some reported mismatches may partly reflect this rather than systematic model biases. Future work with longer storm-resolving simulations will help disentangle sampling effects from structural model errors. 
Also, in regions close to a threshold, classification errors do not necessarily imply large model biases (Section \ref{sec:location_panel_siberia}).

With the wealth of surface observations available today, there is an opportunity to develop a modern classification that incorporates information central to the human experience of climate, such as the diurnal cycle of temperature, the timing and intensity of precipitation, and sunshine hours. The challenge here is to retain an index that is easy to compute from standard model output.

To conclude, while storm-resolving models perform well across many regions, our results also expose areas where local-scale climate representation still falls short of the digital twin ambition. The K\"{o}ppen-Geiger classification, despite being over a century old, remains an intuitive framework for assessing the local quality of global models and for communicating climate to policymakers, ecologists, and the public. If storm-resolving models are to support decision-making in agriculture, water management, and energy, their credibility would increase if they succeeded in reproducing this fundamental climatological framework.

Therefore, we propose that K\"{o}ppen-Geiger maps become a standard diagnostic in storm-resolving model evaluation, complementing process-oriented metrics with a measure of how well global models capture climate as it is experienced locally. This would help bridge the gap between the world of global modelling constrained by the laws of physics and the Humboldtian tradition of observing climate from the ground up.

\section*{Acknowledgements}
This work was supported by the EU Horizon 2020 Project nextGEMS, Grant Agreement Number 101003470. This work used resources of the Deutsches Klimarechenzentrum (DKRZ) granted by its Scientific Steering Committee (WLA) under project IDs bb1153 and bm1235.
LB is funded by the Deutsche Forschungsgemeinschaft (DFG, German Research Foundation) under Germany’s Excellence StrategyEXC 2037 ‘CLICCS—Climate, Climatic Change, and Society’ - Project No. 390683824, a contribution to the 'Earth and Society Research Hub' (ESRAH) of University of Hamburg.
The authors declare no further conflict of interest.

\section*{Data availability}
The data and the scripts to create the figures are available at Zenodo, \url{https://doi.org/10.5281/zenodo.19757734}.
Data availability for the nextGEMS model output data is described in the Data availability section of \cite{Segura2025}.
All data and download instructions of the K\"{o}ppen-Geiger reference data are in \cite{Beck2023}.

\section*{Author contributions}
CvH wrote the manuscript, with input from and review by all authors.
All authors helped shaping the idea during nextGEMS project meetings and hackathons.
CvH and MV designed and performed the analyses.
CvH and CH lead the \textit{Storms and Land} theme.

\printbibliography

\clearpage
\setcounter{figure}{0}
\renewcommand{\thefigure}{S\arabic{figure}}
\section*{Supplement A: K\"{o}ppen-Geiger classification}
\begin{figure}
    \centering
    \includegraphics[width=0.95\linewidth]{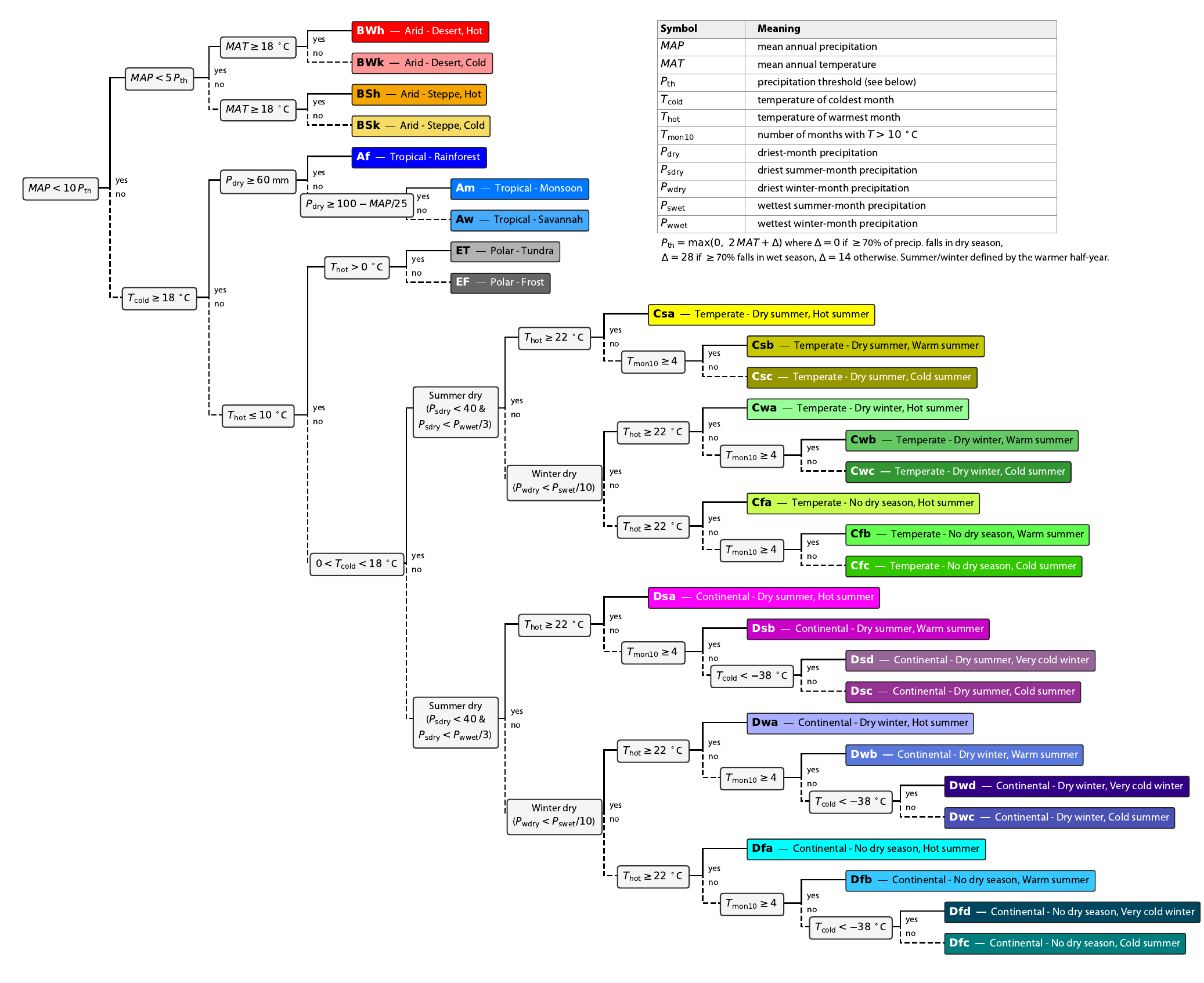}
    \caption{\textbf{K\"{o}ppen-Geiger climate classification decision tree.} The decision tree shows how the classes are computed from monthly mean air temperature and monthly precipitation sums.}\label{fig:koppen_decision_tree}
\end{figure}

Figure \ref{fig:koppen_decision_tree} shows the procedure for deriving the K\"{o}ppen-Geiger classification from monthly mean air temperature and monthly precipitation sums.

\section*{Supplement B: World maps of classification under climate change}
\begin{figure}
    \centering
    \includegraphics[width=0.95\linewidth]{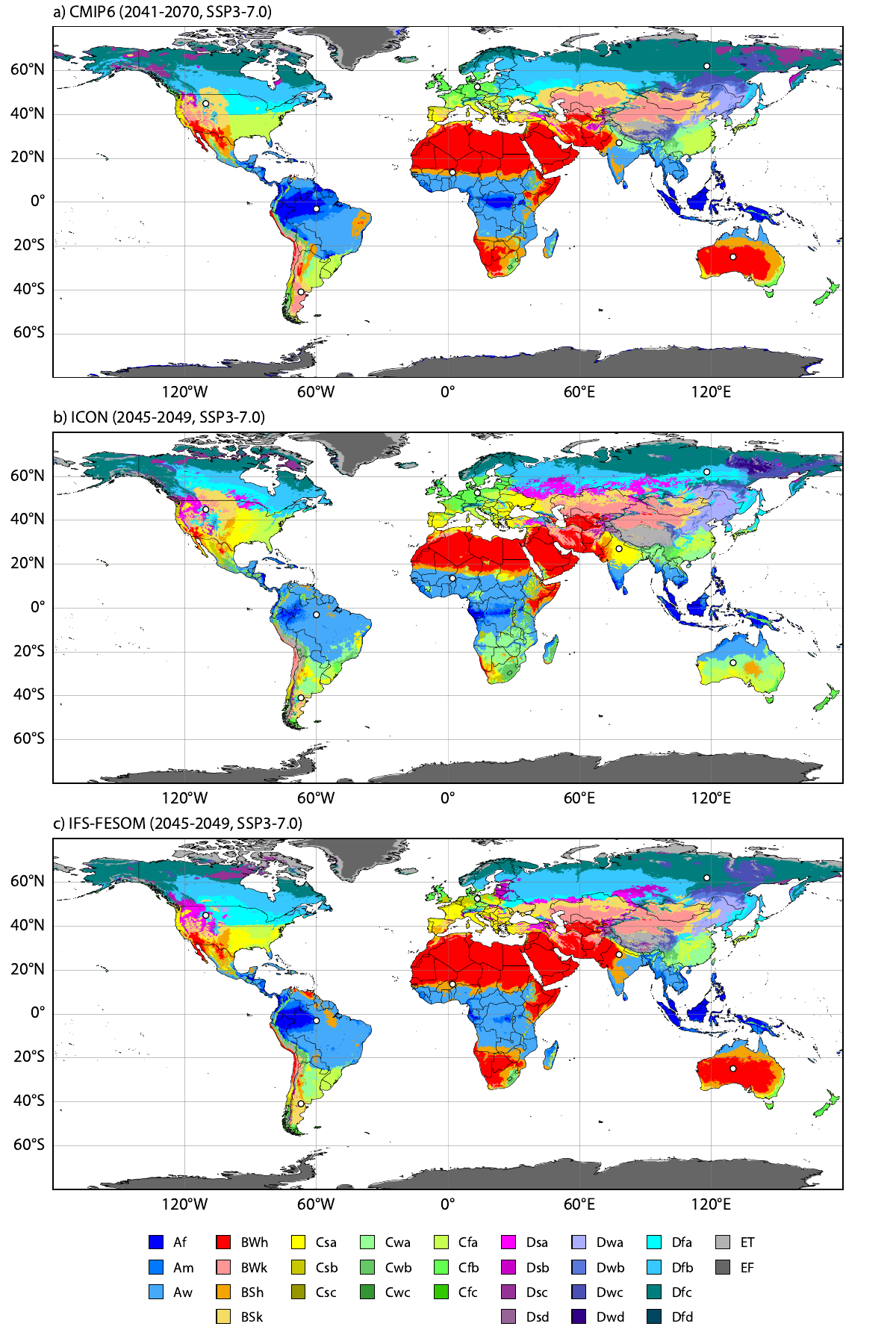}
    \caption{\textbf{K\"{o}ppen-Geiger climate classification for future climate.} Dominant climate zones as derived from \textbf{a)} CMIP6 \cite{Beck2018,Beck2023} and storm-resolving models \textbf{b)}, ICON, and \textbf{c)} IFS-FESOM, as used in Fig. \ref{fig:koppen_bar_charts}. Major climate zones are tropical (A), arid (B), temperate (C), continental (D), and polar (E), see Supplement A for a full overview. Markers indicate the eight locations examined in Sec.~\ref{sec:location_panel}.}
    \label{fig:koppen_future_panels}
\end{figure}

For completion, we show in Fig. \ref{fig:koppen_future_panels} also the world maps with classification under climate change. The data displayed is the same as used in the bar charts in Fig. \ref{fig:koppen_bar_charts}.

\end{document}